\documentclass[fleqn,twoside]{article}
\usepackage[headings]{espcrc2}
\usepackage{a4wide,graphicx,amsmath,axodraw}

\allowdisplaybreaks

\newcommand{\eg}{e.g.\ }
\newcommand{\ie}{i.e.\ }

\newcommand{\rd}{\mathrm{d}}

\newcommand{\uscore}{\symbol{95}}

\title{News from FormCalc and LoopTools}

\author{Thomas Hahn, Michael Rauch%
  \address{Max-Planck-Institut f\"ur Physik \\
  F\"ohringer Ring 6, D--80805 Munich, Germany}
  \hfill MPP-2006-7}

\begin{document}

\begin{abstract}
The FormCalc package automates the computation of FeynArts amplitudes up
to one loop including the generation of a Fortran code for the numerical
evaluation of the squared matrix element.  Major new or enhanced
features in Version 5 are: iterative build-up of essentially arbitrary
phase-spaces including cuts, convolution with density functions, and
uniform treatment of kinematical variables.  The LoopTools library
supplies the one-loop integrals necessary for evaluating the squared
matrix element.  Its most significant extensions in Version 2.2 are the
five-point family of integrals, and complex and alternate versions.
\end{abstract}

\maketitle

%------------------------------------------------------------------------

\section{Introduction}

FormCalc \cite{FormCalc} is a Mathematica package for the calculation of
Feynman amplitudes.  Amplitudes generated by FeynArts \cite{FeynArts}
are simplified analytically and converted to a self-contained Fortran 
code for the computation of the squared matrix element.  The present 
article describes the following features added or enhanced in Version 5:
\begin{itemize}
\item Iterative build-up of essentially arbitrary phase-spaces 
      with various cuts,

\item Convolution with density functions for external particles,

\item Uniform treatment of kinematic variables,

\item Better modularity and code reusability, \ie less cross-talk 
      between program modules.
\end{itemize}
The LoopTools library supplies the implementations of the scalar and
tensor one-loop integrals necessary for running the Fortran code 
generated by FormCalc.  The extensions in Version 2.2 are:
\begin{itemize}
\item Addition of the five-point family of functions,

\item A collective two-point function \texttt{Bget} similar to 
      \texttt{Cget} and \texttt{Dget},

\item Complex versions of most integrals,

\item Run-time selection of alternate versions for checking,

\item A command-line interface,

\item Enhanced cache functionality,

\item Better internal accuracy, and a complete quadruple precision
      build available on Intel x86/ifort platforms.
\end{itemize}

%------------------------------------------------------------------------

\section{FormCalc}

\subsection{Phase-space}

FormCalc 5 no longer uses hand-tailored phase-space parameterizations
like its predecessor versions, but builds up the $n$-particle
phase-space iteratively by nested integrations over the invariant mass 
$M_i$ and solid angle $\Omega_i$ of each outgoing particle $i$.  This 
procedure is encoded in the subroutine \texttt{Split}:
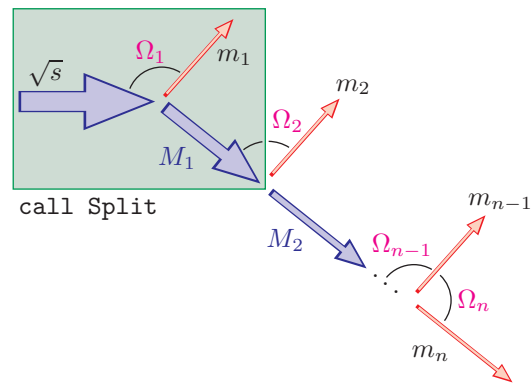
\begin{center}
\begin{picture}(187,143)(-2,-5)
\CBox(-2,67)(93,135){ForestGreen}{PastelGreen}
\Text(0,64)[lt]{\texttt{call Split}}

\SetWidth{8}
\SetColor{Blue}
\LongArrow(0,100)(50,100)
\SetWidth{7}
\SetColor{PastelBlue}
\LongArrow(1,100)(48,100)
\Text(10,106)[b]{$\sqrt s$}

\SetWidth{1.5}
\SetColor{Red}
\LongArrow(55,102)(80,130)
\SetWidth{.9}
\SetColor{PastelRed}
\LongArrow(55.8,103)(79,129)
\Text(75,120)[lt]{$m_1$}

\SetWidth{.5}
\SetColor{Black}
\CArc(54,100)(13,58,158)
\Text(50,115)[b]{$\Magenta{\Omega_1}$}

\SetWidth{5}
\SetColor{Blue}
\LongArrow(55,98)(90,70)
\SetWidth{4}
\SetColor{PastelBlue}
\LongArrow(55.8,97.3)(88.5,71.3)
\Text(67,83)[tr]{\Blue{$M_1$}}

\SetWidth{1.5}
\SetColor{Red}
\LongArrow(95,72)(120,100)
\SetWidth{.9}
\SetColor{PastelRed}
\LongArrow(95.8,73)(119,99)
\Text(120,103)[lb]{$m_2$}

\SetWidth{.5}
\SetColor{Black}
\CArc(94,70)(15,58,89)
\CArc(94,70)(15,98,128)
\Text(96,93)[l]{$\Magenta{\Omega_2}$}

\SetWidth{3}
\SetColor{Blue}
\LongArrow(95,66)(130,38)
\SetWidth{2}
\SetColor{PastelBlue}
\LongArrow(95.8,65.3)(128.5,39.3)
\Text(108,52)[tr]{\Blue{$M_2$}}

\Text(133,42)[lt]{$\ddots$}

\SetWidth{1.5}
\SetColor{Red}
\LongArrow(150,28)(175,56)
\SetWidth{.9}
\SetColor{PastelRed}
\LongArrow(150.8,29)(174,55)
\Text(170,59)[lb]{$m_{n-1}$}

\SetWidth{.5}
\SetColor{Black}
\CArc(149,25)(13.5,60,140)
\Text(145,41)[b]{$\Magenta{\Omega_{n-1}}$}
\CArc(150,25)(13,-36,47)
\Text(166,25)[l]{$\Magenta{\Omega_n}$}

\SetWidth{1.5}
\SetColor{Red}
\LongArrow(150,23)(185,-5)
\SetWidth{.9}
\SetColor{PastelRed}
\LongArrow(150.8,22.3)(184,-4.1)
\Text(163,8)[tr]{$m_n$}
\end{picture}
\end{center}
Counting the degrees of freedom, there are $(n - 1)$ $M$-integrations 
and $n$ $\Omega$-integrations.  The corresponding phase-space 
parameterization is
\begin{align}
&\frac 1{2\sqrt s}\int_{m_2+\dots+m_n}^{\sqrt s - m_1}
  \rd M_1\,\rd\Omega_1\,\frac{k_1}{2} \notag \\
&\times\int_{m_3+\dots+m_n}^{M_1 - m_2}
  \rd M_2\,\rd\Omega_2\,\frac{k_2}{2} \notag \\
&\times\cdots \notag \\
&\times\int_{m_n}^{M_{n-2} - m_{n-1}}
  \rd M_{n-1}\,\rd\Omega_{n-1}\,\frac{k_{n-1}}{2} \notag \\
&\times\int
  \rd\Omega_n\,\frac{k_n}{2}
\end{align}
where $\rd\Omega_i = \rd\cos\theta_i\:\rd\varphi_i$.  The particle's
momentum $k_i$ and $\cos\theta_i$ are given in the respective decay's
rest frame.  The $\varphi_1$-integration is trivial because of axial 
symmetry.  From the practical point of view this looks as follows
(this code is taken almost verbatim from FormCalc's \texttt{2to3.F}):

\begin{small}
\begin{verbatim}
p = 0
ex = 0
ey = 0  
ez = 1
minv = sqrtS
msum = MASS3 + MASS4 + MASS5

call Split(5, dble(MASS5),
& p, ex,ey,ez, minv, msum, fac, 0, 
& Var(XMREM5), Var(XCOSTH5), Var(TRIVIAL))

call Split(4, dble(MASS4),
& p, ex,ey,ez, minv, msum, fac, 0,
& Var(FIXED), Var(XCOSTH4), Var(XPHI4))

call VecSet(3, dble(MASS3), p, ex,ey,ez)
\end{verbatim}
\end{small}
One starts with the initial reference direction in $(\mathtt{ex},
\mathtt{ey}, \mathtt{ez})$ and no boost, \texttt{p = 0}.  The available
energy is given in \texttt{minv} and the sum of external masses in
\texttt{msum}. The \texttt{Split} subroutine is then called $(n - 1)$
times for an $n$-particle final state.  The reference direction, the
boost, \texttt{minv}, and \texttt{msum} are automatically adjusted along
the way for the respective remaining subsystem and ultimately determine
the remaining $n$-th vector unambigously, which is then simply set by
\texttt{VecSet}.

About the integration variables more will be said in the next section. 
For the moment, note that the \texttt{X} in \texttt{XMREM5} refers to
the ratio, \ie \texttt{XMREM5} runs from 0 to 1.  The actual integration
borders are determined internally by Split.

%------------------------------------------------------------------------

\subsection{Variables}

FormCalc 5 introduces a new homogeneous system for all (potential)
integration variables.  Each variable is referred to by a preprocessor
constant, \eg {\tt SQRTS} or {\tt XCOSTH}.  The following parts 
can be accessed via preprocessor macros:
\begin{itemize}
\item
\texttt{Var($v$)} = the actual value of $v$.

\item
\texttt{Show($v$)} = the value printed in the output
-- to print \eg $t$ instead of $\cos\theta$.

\item
\texttt{Lower($v$)}, \texttt{Upper($v$)}, \texttt{Step($v$)} =
the lower limit, upper limit, and step width of $v$.
If the step is zero, the cross-section is integrated over $v$.

\item
\texttt{CutMin($v$)}, \texttt{CutMax($v$)} = the lower and upper cuts 
on $v$.
\end{itemize}
There are two special variables: \texttt{FIXED} for fixed values,
\ie no integration, and \texttt{TRIVIAL} for trivial integrations.

%------------------------------------------------------------------------

\subsection{Cuts}

Split allows to place cuts on each $M$- and $\cos\theta$-integration.  
The $\varphi$-integration is not modified in the present setup.
\begin{center}
\begin{tabular}{|c|l|}
\multicolumn{2}{c}{Cuts restricting $M_i$} \\ \hline
Cut on & Key \\ \hline
$M_i$     & \verb=CUT_MREM=      \\
$E_i$     & \verb=CUT_MREM_E=    \\
$k_i$     & \verb=CUT_MREM_K=    \\
$E_{T,i}$ & \verb=CUT_MREM_ET=   \\
$k_{T,i}$ & \verb=CUT_MREM_KT=   \\
$y_i$     & \verb=CUT_MREM_RAP=  \\
$\eta_i$  & \verb=CUT_MREM_PRAP= \\ \hline
\multicolumn{2}{c}{} \\
\multicolumn{2}{c}{Cuts restricting $\cos\theta_i$} \\ \hline
Cut on & Key \\ \hline
$\cos\theta_i$     & \verb=CUT_COSTH=    \\
$\cos\hat\theta_i$ & \verb=CUT_COSTHCMS= \\
$E_i$              & \verb=CUT_COSTH_E=  \\
$k_i$              & \verb=CUT_COSTH_K=  \\ \hline
\end{tabular}
\end{center}
In practice, the application of cuts works as follows:
\begin{small}
\begin{verbatim}
key = 0

CutMin(XMREM5) = E5MIN
key = key + Cut(CUT_MREM_E, CUT_MIN)

CutMin(XCOSTH5) = -(1 - COSTH5CUT)
CutMax(XCOSTH5) = +(1 - COSTH5CUT)
key = key +
& Cut(CUT_COSTH, CUT_MIN + CUT_MAX)

call Split(5, dble(MASS5),
& p, ex,ey,ez, minv, msum, fac, key,
& Var(XMREM5), Var(XCOSTH5), Var(TRIVIAL))
...
\end{verbatim}
\end{small}
The value of the cut is deposited in \texttt{CutMin} or \texttt{CutMax}
and the cut registered by adding an identifier for the cut to the
integer \texttt{key}, \eg \texttt{Cut(CUT\uscore MREM\uscore E,
CUT\uscore MIN)} specifies a cut on the energy (\texttt{CUT\uscore
MREM\uscore \textit{E}}) from below (\texttt{CUT\uscore \textit{MIN}})
which is used to restrict the invariant-mass integration
(\texttt{CUT\uscore \textit{MREM}\uscore E}).

Note that these cuts really restrict the integration limits.  They do 
not introduce veto functions (1 in wanted, 0 in unwanted areas) into the 
integrand, which can severely hamper convergence.

In addition, for each external particle $i$ several kinematical 
quantities are available:
\begin{itemize}
\item \texttt{momspec(SPEC\uscore M,$i$)} \\
      -- mass $m$,

\item \texttt{momspec(SPEC\uscore E,$i$)} \\
      -- energy $E$,

\item \texttt{momspec(SPEC\uscore K,$i$)} \\
      -- momentum $k$,

\item \texttt{momspec(SPEC\uscore ET,$i$)} \\
      -- transverse energy $E_T$,

\item \texttt{momspec(SPEC\uscore KT,$i$)} \\
      -- transverse momentum $k_T$,

\item \texttt{momspec(SPEC\uscore RAP,$i$)} \\
      -- rapidity $y$,

\item \texttt{momspec(SPEC\uscore PRAP,$i$)} \\
      -- pseudo-rapidity $\eta$,

\item \texttt{momspec(SPEC\uscore DELTAK,$i$)} \\
      -- the difference $E - k$.
\end{itemize}

%------------------------------------------------------------------------

\subsection{Convolution}

With the new variable system, the convolution with arbitrary parton
distribution functions can easily be achieved.  Three modules are
already included in FormCalc 5:
\begin{itemize}
\item \texttt{lumi\uscore parton.F}
      = initial-state partons, no convolution.

\item \texttt{lumi\uscore hadron.F}
      = initial-state hadrons,
      convolution with hadronic PDFs from the LHAPDF library 
      \cite{LHAPDF}.

\item \texttt{lumi\uscore photon.F}
      = initial-state photons,
      convolution with CompAZ spectrum \cite{CompAZ}.
\end{itemize}

%------------------------------------------------------------------------

\subsection{Modularity and Code-Reusability}

The choice of parameters is directed as in previous versions by the two 
files \texttt{process.h} and \texttt{run.F}, which include one each of
\begin{itemize}
\item Kinematics definitions: \\
      \texttt{1to2.F}, \\
      \texttt{2to2.F}, \\
      \texttt{2to3.F},

\item Convolution with PDFs: \\
      \texttt{lumi\uscore parton.F}, \\
      \texttt{lumi\uscore hadron.F}, \\
      \texttt{lumi\uscore photon.F}, 

\item Model initialization: \\
      \texttt{model\uscore sm.F}, \\
      \texttt{model\uscore mssm.F}, \\
      \texttt{model\uscore thdm.F}.
\end{itemize}
There is almost no cross-talk between different modules which are in
that sense `universal.'  Also, the main program has been radically 
slimmed down in FormCalc 5.  It now only scans the command line and 
invokes
\begin{verbatim}
call ProcessIni(...)
call ParameterScan(...)
\end{verbatim}
All further action is decoupled from the main program and can easily be
called from any application.  It is thus straightforward to use
Form\-Calc-generated code in own programs.

%------------------------------------------------------------------------

\section{LoopTools}

LoopTools is a library for the one-loop integrals.  It is based on FF
\cite{FF} and has a Fortran, C/C++, and Mathematica interface.  It is
referenced by the FormCalc-generated code but can of course be used
also without FormCalc.

%------------------------------------------------------------------------

\subsection{Five-point functions}

The most significant addition is the five-point family of functions --
the scalar integral $E_0$ and the tensor coefficients up to rank four. 
The default versions use the Denner--Dittmaier decomposition \cite{e0}
in which one inverse power of the Gram determinant is cancelled and thus
has considerably better numerical stability than the Passarino--Veltman
\cite{pave} decomposition, which is also available for comparison.

%------------------------------------------------------------------------

\subsection{Combined two-point function}

The two-point functions have been united into the \texttt{Bget} function
which works similar to its \texttt{Cget}, \texttt{Dget}, and
\texttt{Eget} counterparts, in particular it caches its results.
Compatibility routines for the old \texttt{B0}, \texttt{B1}, etc.\ are
of course available.  The reason is mainly cache efficiency in view of 
the five-point decomposition:
\begin{align}
\mathtt{Eget}
\overset{\text{calls}}{\longrightarrow}
& \:5\,\mathtt{Dget} \notag \\
\overset{\text{call}}{\longrightarrow}
& \:(5\cdot 4)\,\mathtt{Cget} \notag \\
\overset{\text{call}}{\longrightarrow}
& \:(5\cdot 4\cdot 3)\,\mathtt{Bget}
\end{align}

%------------------------------------------------------------------------

\subsection{Complex versions}

Versions of the LoopTools functions for complex parameters have been
added as far as they are contained in FF, \ie currently only some
special cases for the complex $D_0$ are available.
\begin{center}
\begin{tabular}{l|l}
real versions & complex versions \\ \hline
\texttt{A0}, \texttt{A00} &
  \texttt{A0C}, \texttt{A00C} \\
\texttt{B0}, \texttt{B1}, \dots &
  \texttt{B0C}, \texttt{B1C}, \dots \\
\texttt{B0i}, \texttt{Bget} &
  \texttt{B0iC}, \texttt{BgetC} \\
\texttt{C0}, \texttt{C0i}, \texttt{Cget} &
  \texttt{C0C}, \texttt{C0iC}, \texttt{CgetC} \\
\texttt{D0}, \texttt{D0i}, \texttt{Dget} &
  \texttt{D0C}, \texttt{D0iC}, \texttt{DgetC} \\
\texttt{E0}, \texttt{E0i}, \texttt{Eget} &
  \texttt{E0C}, \texttt{E0iC}, \texttt{EgetC}
\end{tabular}
\end{center}

%------------------------------------------------------------------------

\subsection{Alternate versions}

For some functions alternate versions exist, most of which are based on
an implementation by Denner.  The user can choose at run-time whether
the default version `a' (mostly FF) or the alternate version `b' (mostly
Denner) is used and whether checking is performed.  This is determined
by the version key:
\begin{tabbing}
\texttt{0*key}~ \= compute version `a', \\
\texttt{1*key} \> compute version `b', \\
\texttt{2*key} \> compute both, compare, return `a', \\
\texttt{3*key} \> compute both, compare, return `b'.
\end{tabbing}
Usage is as in
\begin{verbatim}
call setversionkey(2*KeyC0 + 3*KeyD0)
\end{verbatim}
The following keys for alternate versions are currently available: 
\texttt{KeyA0}, \texttt{KeyBget}, \texttt{KeyC0}, \texttt{KeyD0}, 
\texttt{KeyEget}, \texttt{KeyEgetC}.  \texttt{KeyAll} comprises all of 
these.

%------------------------------------------------------------------------

\subsection{Debugging}

Debugging output can be turned on similarly with \eg
\begin{verbatim}
call setdebugkey(DebugC + DebugD)
\end{verbatim}
Identifiers range from \texttt{DebugB} to \texttt{DebugE} and are 
summarized by \texttt{DebugAll}.  The integrals are listed in the output 
with a unique serial number.  If the list of integrals becomes too 
long, one can select only a range of serial numbers for viewing, as in
\begin{verbatim}
call setdebugrange(4711, 4715)
\end{verbatim}
This makes it easy to monitor `suspicious' integrals.

%------------------------------------------------------------------------

\subsection{Environment variables}

Version key, debug key, and debug range can be set also from the
`outside,' \ie through environment variables, thus making it unnecessary 
to recompile the code.  Unfortunately, the \texttt{Key\textit{X}} 
and \texttt{Debug\textit{X}} labels cannot be used on the shell 
command-line, but \texttt{-1} offers a convenient way to set all bits 
of an integer, thus having the same effect as \texttt{KeyAll} or 
\texttt{DebugAll}.  For example:
\begin{verbatim}
setenv LTVERSION -1
setenv LTDEBUG -1
setenv LTRANGE 4711-4715
\end{verbatim}

%------------------------------------------------------------------------

\subsection{Command-line Interface}

The new command-line interface is useful in particular for testing and
debugging.  It lists the $N$-point scalar and tensor coefficients
corresponding to the number of arguments, \ie 3 arguments = B, 6
arguments = C, etc.  Currently only the real versions are accessible
through the command-line interface, mainly because it is not
straightforward syntactically to specify complex parameters on the
command-line.

%------------------------------------------------------------------------

\subsection{LoopTools caches}

The internal caching mechanism is meanwhile used by the following
functions: \texttt{Bget}, \texttt{Cget}, \texttt{Dget}, \texttt{Eget},
\texttt{BgetC}, \texttt{CgetC}, \texttt{DgetC}, and \texttt{EgetC}. 
Obviously the former system with its \texttt{getcachelast} and
\texttt{setcachelast} call for each individual cache was no longer 
practicable for all those caches.

The new cache-management functions operate on all caches simultaneously:
\begin{itemize}
\item \texttt{call clearcache} \\
      -- clears all caches,
\item \texttt{call markcache} \\
      -- marks the current position,
\item \texttt{call restorecache} \\
      -- reverts to the last marked position.
\end{itemize}
For compatibility, LoopTools still includes two routines 
\texttt{getcachelast} and \texttt{setcachelast}.  They work only 
approximately as before, however, and are therefore deprecated.

Furthermore, the cache mechanism itself has been improved and now uses a
binary search method.  Cache lookups are thus now faster, but as this
has not often been a bottleneck, the impact on program performance will
typically be minor in most applications.

%------------------------------------------------------------------------

\subsection{Accuracy}

The accuracy of the tensor reduction has been improved through the use
of an LU decomposition for the Gram matrix and quadruple precision at
strategic points internally.  Thanks to the Intel ifort compiler, the
latter is now widely available on the x86 platform.

A complete quadruple-precision build can be chosen with an alternate
makefile.  In particular the quadruple-precision version of the frontend
can be useful to check cases where a severe loss of precision is
suspected.

%------------------------------------------------------------------------

\section{Summary}

FormCalc 5 is the current release of the FormCalc package with the 
following new features and extensions:
\begin{itemize}
\item New kinematics routines can build up almost arbitrary
      phase-spaces with a wide range of cuts possible.
\item FormCalc's driver code consists of intuitively programmed,
      concise modules with almost no cross-talk between them.  
      Calling FormCalc-generated code from other applications
      has been much simplified.
\item Convolution with arbitrary distribution functions is possible.  
      The following common cases are available out-of-the-box
      (if FormCalc came in one):
      \begin{itemize}
      \item partons (no convolution),
      \item hadrons (uses LHAPDF),
      \item photons (uses CompAZ).
      \end{itemize}
\item All possible integration variables are treated uniformly and
      detailed kinematical information is available for all external 
      legs.
\end{itemize}
LoopTools 2.2 is the latest version of the LoopTools library with
the following novelties:
\begin{itemize}
\item The scalar five-point function including its tensor coefficients
      up to rank four is provided.
\item Complex versions are available as far as implemented in FF.
\item Checking can be enabled at run-time, individually for all
      integrals which have alternate versions, by program call or
      through an environment variable.
\item In the same manner, debugging output can be turned on 
      and off individually at run-time.
\item The cache system has been extended and improved.
\item A new command-line interface is useful for testing and debugging.
\item The accuracy of the tensor reduction has been enhanced.
\item A complete quadruple-precision build can be chosen with an
      alternate makefile (\eg \texttt{make -f makefile.quad-ifort}).
\end{itemize}
Both packages are available as open source and stand under the GNU 
Lesser General Public License (LGPL).  They can be obtained from the 
Web sites \\
\texttt{http://www.feynarts.de/formcalc}, \\
\texttt{http://www.feynarts.de/looptools}.

%------------------------------------------------------------------------

\section*{Acknowledgments}

We thank S.~Dittmaier for significant help with the implementation and 
comparison of the five-point functions and C.~Schappacher for intensive 
beta-testing.

%------------------------------------------------------------------------

\begin{flushleft}

\end{flushleft}


\begin{thebibliography}{99}

\bibitem{FormCalc}
T.~Hahn, M.~P\'erez-Victoria, \textsl{Comp.\ Phys.\ Commun.}
\textbf{118} (1999) 153 [hep-ph/9807565].
  %%CITATION = HEP-PH 9807565;%%

\bibitem{FeynArts}
T.~Hahn, \textsl{Comp.\ Phys.\ Commun.} \textbf{140} (2001) 418
[hep-ph/0012260].
  %%CITATION = HEP-PH 0012260;%%

\bibitem{LHAPDF}
M.R.~Whalley, D.~Bourilkov, R.C.~Group, hep-ph/0508110,
  %%CITATION = HEP-PH 0508110;%%
http://hepforge.cedar.ac.uk/lhapdf.

\bibitem{CompAZ}
A.F.~Zarnecki, \textsl{Acta Phys.\ Polon.} \textbf{B34} (2003) 
2741 [hep-ex/0207021].
  %%CITATION = HEP-EX 0207021;%%

\bibitem{FF}
G.J.~van Oldenborgh, J.A.M.~Vermaseren, \textsl{Z.\ Phys.\ C} 
\textbf{46} (1990) 425.
  %%CITATION = ZEPYA,C46,425;%%

\bibitem{e0}
A.~Denner, S.~Dittmaier, \textsl{Nucl.\ Phys.\ B} 658 (2003) 175
[hep-ph/0212295].
  %%CITATION = HEP-PH 0212259;%%

\bibitem{pave}
G.~Passarino, M.J.~Veltman, \textsl{Nucl.\ Phys.\ B} \textbf{160} 
(1979) 151.
\end{thebibliography}
\end{document}